\documentclass{article}
\usepackage{spconf,amsmath,graphicx}
\usepackage{amsfonts,amsmath}
\usepackage{xcolor}
\usepackage{graphicx}
\usepackage{cite}
\usepackage[title]{appendix}
\usepackage[utf8]{inputenc}
\usepackage{enumitem}
\usepackage[linesnumbered,algoruled,boxed,lined]{algorithm2e}

\graphicspath{{figures/}}

\def\e{\mathrm{e}}
\def\E{\mathrm{E}}

\newcommand{\argmax}{\operatornamewithlimits{argmax}}

\newcommand{\tcr}[1]{#1}

\usepackage{pifont}
\newcommand{\cm}{\ding{51}}%
\let\oldbibliography\thebibliography
\renewcommand{\thebibliography}[1]{%
  \oldbibliography{#1}%
  \setlength{\itemsep}{1pt}%
}

\title{GENERATIVE SPEECH CODING WITH PREDICTIVE VARIANCE REGULARIZATION} 
\name{\begin{tabular}{c}
   W. Bastiaan Kleijn,$^{1,2}$ 
  Andrew Storus,$^{1}$
 Michael Chinen,$^{1}$
 Tom Denton,$^{1}$ \\
 Felicia S. C. Lim,$^{1}$
 Alejandro Luebs,$^{1}$
 Jan Skoglund,$^{1}$
 Hengchin Yeh$^{1}$
 \vspace{-1mm}
  \end{tabular}}
\address{$^1$ Google LLC, San Francisco, USA\\
         $^2$ School of Engineering and Computer Science, Victoria University of Wellington, New Zealand    
\vspace{-1mm}
}
\begin{document}
\ninept

\maketitle

\begin{abstract}
The recent emergence of machine-learning based generative models for speech
suggests a significant reduction in bit rate for speech codecs is
possible. However, the performance of generative models deteriorates
significantly with the distortions present in real-world input signals. We argue
that this deterioration 
is due to the sensitivity of the maximum likelihood criterion to outliers and
the ineffectiveness of modeling a sum of independent signals with a single
autoregressive model. We introduce predictive-variance regularization to reduce
the sensitivity to outliers, resulting in a significant increase in performance. We
show that noise reduction to remove unwanted signals can significantly
increase performance. We provide extensive subjective performance evaluations
that show that our system based on generative modeling provides state-of-the-art coding
performance at 3 kb/s for real-world speech signals at reasonable computational complexity.
\end{abstract}

\section{Introduction}
\label{s:intro}

In recent years it has become possible to generate high-quality speech from a
conditioning sequence with a very low information rate. This suggests that
generative synthesis forms a natural basis for the coding and enhancement of
speech. However, it has been found that generative synthesis is sensitive to the
quality of the data used for training and the conditioning sequences used for
training and inference. This can result in poor synthesized speech quality. In this
paper, we discuss methods that significantly reduce the impact of distortions in
the input signal on signal synthesis for speech coding. 

WaveNet \cite{oord2016wavenet} first showed that the
generation of high-quality speech from 
only a low-rate conditioning sequence, such as written text, is possible. WaveNet is
based on an autoregressive structure that specifies a predictive distribution
for each subsequent signal sample.  While WaveNet uses a dilated convolution to
determine the predictive distribution, other recurrent neural networks structures such
as WaveRNN \cite{kalchbrenner2018efficient} and the WaveGRU structure that we
use in this paper have also been used successfully for this purpose. Although 
autoregressive structures for synthesis are common,  
feed-forward structures are used by, for example, Parallel WaveNet
\cite{oord2018parallel}, WaveGlow \cite{prenger2019waveglow}, WaveGAN \cite{donahue2018adversarial},
and GANSynth \cite{engel2019gansynth}. It is fair to state that while the more
recent methods may have computational advantages, they do not surpass the basic
synthesis quality of the original WaveNet approach. 

The high quality of generative speech synthesis has led to a significant effort
towards its usage for coding.  In contrast to synthesis from text, synthesis for
coding must be able to generate an unlimited range 
of voices and its conditioning is variable as it is computed from input signals
that may suffer from a range of distortions. It was found that the synthesis
of a wide range of voices with a single generative model is not a significant
problem. Generative synthesis of a wide range of unknown voices with a single
model results in only a minor reduction of speaker identifiability
\cite{kleijn2018wavenet}. However, variability in the conditioning leads to a reduced
output speech quality that can not be improved significantly with
straightforward measures such as training with noisy conditioning and
undistorted target signals. Hence, despite \tcr{extensive research},
e.g., \cite{klejsa2019high,garbacea2019low,valin2019real,lim2020robust,fejgin2020source} generative synthesis based speech coding has not yet seen major practical
applications. 

The contribution of this paper consists of the identification of causes of
the sensitivity to distortion, the development of methods to reduce this
sensitivity, and subjective testing of the new methods confirming the
improvements. We show that a major cause of the sensitivity is associated
with an attribute of the log-likelihood (LL) objective function. The LL
objective function incurs a high penalty if the model assigns a low probability
to observed data. Hence, in the context of autoregressive structures, it encourages
an overly broad predictive distribution when at least some training  
data are difficult to predict accurately from the past signal and conditioning, which is true for
real-world training data. We mitigate this effect by including predictive variance
regularization in the objective function.

\tcr{We also show with experiments that input-noise
suppression can improve performance significantly. It is well-known that a
sum of low-order linear autoregressive processes is, in general, not a low-order
autoregressive process. This suggests that linear autoregressive models are poor
for sums of independent signals and our results indicate this also
holds for nonlinear autoregressive models.  Whereas traditional
analysis-by-synthesis coding methods can compensate for model inadequacies,
this is not true for generative synthesis based coding, explaining the
effectiveness of noise suppression.} 


\section{Problem formulation}
\label{s:Problem}

In this section, we first describe how an autoregressive model is used to
model a process. The method is as proposed in \cite{oord2016wavenet}.  We then
discuss a common problem that occurs when training such sequences.

Consider a random process $\{X_i\}$ consisting of real-valued random samples
$X_i$, with a  time
index $i \in \mathbb{Z}$. The joint distribution of a finite sequence, $p(x_{i},\cdots, x_{i-N})$,
can be expressed as a product of conditional distributions:
\begin{align}
p(x_{i},\cdots, x_{i-N}|\beta) =  \prod_{j=0}^{N} p(x_{i-j} | x_{i-j-1},\cdots, x_{i-N}, \beta), \label{q:conditional}
\end{align}
where $\beta$ is conditioning information.

It follows from \eqref{q:conditional} that we can create an approximate
realization of a random  process by recursively sampling from a model of the
predictive distribution $p(x_i | x_{i-1},\cdots, x_{i-N},\beta)$ for sufficiently
large $N$. It is convenient to use a standard-form distribution $q(x_i |
\alpha)$ with parameters $\alpha$ as a model predictive distribution. The
standard-form distribution can be a Gaussian or a logistic mixture, for
example. This formulation allows us to predict the model parameters with a 
deterministic neural network $\phi:  (x_{i-1}, \cdots, x_{i-N},\beta,W ) \mapsto
\alpha$ where $W$ is a vector of network parameters. Thus, the predictive
distribution for sample $x_i$ is now $q(x_i| \phi (x_{i-1}, \cdots, x_{i-N},
\beta, W))$.

To find the parameters $W$, a reasonable objective is to minimize the
Kullback-Leibler divergence between the ground truth joint distribution $p(x_i,\cdots,
x_{i-N})$ and the model distribution $q(x_i,\cdots,x_{i-N})$, or, equivalently,
the cross-entropy between these distributions. The latter measure is tractable even
though $p$ is only available as an empirical distribution. It follows
from \eqref{q:conditional} and our formulation of $q(x_i | \alpha)$ 
that cross-entropy based estimation of the parameters of $\phi$ can be
implemented using maximum-likelihood based teacher forcing. For a database of
$M$ signal samples, the maximum-likelihood estimate of $W$ can be written as  
\begin{align}
W^*  =  \argmax_W \sum_{i=1}^M \log q(x_{i} |\phi( x_{i-1},\cdots,  x_{i-N}, 
\beta, W)). \label{q:MLobj0}
\end{align}
Note that \eqref{q:MLobj0} leads to rapid training as it facilitates parallel implementation.
For sufficiently large $N$ and $M$, the LL objective provides an upper bound on
the differential entropy rate as
\begin{multline}
h(X_i |X_{i-j}, \cdots, X_{i-N}) \leq \\-\frac{1}{M}\sum_{i=1}^M \log q(x_{i}
  |\phi( x_{i-1},\cdots,  x_{i-N},  W)), \label{q:condEnt}
\end{multline}
where, for notational convenience, we considered the unconditioned case.
Conversely, \eqref{q:condEnt} can be interpreted as a lower bound on a measure of
uncertainty associated with the model predictive distribution. This lower bound
is associated with the process itself and not with the model.

Although the differential entropy rate is subadditive for summed signals, predictive models tend not to work well for summed signals.  
In general, a model of summed signals is essentially multiplicative in the required model configurations. It is well-known that the sum of 
finite-order linear autoregressive models is, in general, not a finite-order
autoregressive model \cite{granger1976time}. It is relatively straightforward to reduce this problem with noise suppression.

A more difficult problem relates to well-known drawbacks of the Kullback-Leibler divergence and, hence, the LL
objective of \eqref{q:MLobj0}. When the model
distribution $q$ vanishes in the support region of the groundtruth $p$, the
Kullback-Leibler divergence diverges.  In \eqref{q:MLobj0} this manifests itself
as a severe penalty for training data $x_i$ that have a low model probability $q(x_{i}
|\phi( x_{i-1},\cdots,  x_{i-N},  \beta, W))$. Hence, a few nonrepresentative outliers
in the training data may lead the training procedure to equip
the predictive model distribution with heavy tails. Such tails lead to
signal synthesis with a relatively high entropy rate during inference.  In audio
synthesis this corresponds to a noisy synthesized signal. Hence it is
desirable to counter the severity of the penalty for low probability training data.

We can identify a second relevant drawback to the ML objective. When the ML objective function is used, the
model distribution should converge to the 
groundtruth distribution with increasing database size. However, in practice the
stochastic nature of the training data and the training method results in
inaccuracies and this in turn means the method attempts to minimize the impact of such errors.  For example, the implicit description of pitch by the  
predictive distribution may be inaccurate.
A predictive model distribution with heavy tails 
for voiced speech then increases the likelihood
of training data as it reduces the impact of the model pitch deviating from the 
groundtruth pitch. From this reasoning we conclude that it is desirable to
account for the audibility (perception) of distortions, leading to empirically
motivated refinements of the objective function.  

The problems associated with the LL objective have been considered earlier in different contexts. The
vanishing support problem described above was addressed in the context of 
generative adversarial networks (GANs) \cite{goodfellow2014generative}, where the implicit Jensen-Shannon
objective function of the original method and the more general $ f $-divergence
based method \cite{nowozin2016f} suffer, at least in principle, from similar
problems. The support problem in GANs can be removed by using the 1-Wasserstein
distance \cite{arjovsky2017wasserstein} or with maximum mean discrepancy (MMD)
\cite{li2015generative,li2017mmd}. However, as these measures require as input two
empirical distributions, these methods are natural for static distributions and
not for dynamic predictive distributions. The methods also do not 
facilitate adjustment to account for perception. An existing approach that
attempts to compensate for overly broad predictive distributions is to lower the
``temperature" during inference, e.g., \cite{kim2018flowavenet}.  The predictive distribution is typically
raised to a power, then renormalized. This approach does not account for the implicit cost penalty in the basic training objective.

\section{Objective functions for predictive distribution models}
\label{s:objective}

In this section, we discuss two related approaches that modify the maximum likelihood criterion
to obtain improved performance. Both approaches aim to reduce the impact of data 
points in the training set that are difficult to predict. The methods remove the
need for heuristic modifications during inference. While the principles of our methods are general, we apply them to the mixture of logistics distribution that we use in our coding scheme (cf. section \ref{s:system}).

\subsection{Encouragement of low-variance predictive distributions}
\label{s:direct}
We now discuss how to add a term to the objective function that encourages
low-variance predictive distributions. In this approach we define the overall objective
function for the weights $W$ given a database $\{x\}$ as
\begin{align}
  J(\{x\}, W) = J_{\mathrm{LL}}( \{x\}; W) + \nu J_{\mathrm{var}}(\{x\},W) \label{q:genObjFunc}
\end{align}
where the log likelihood over the database, $J_{\mathrm{LL}}( \{x\}; W)= \E_{\mathrm{data}} \log q(x_{i} ;\phi(
x_{i-1},\cdots,  x_{i-N}, W))$,  is combined with a variance regulatization term  
$J_{\mathrm{var}}(\{x\},W)$ that is defined
below and where $\nu$ is
a constant that must be tuned.

\subsubsection{Predictive variance computation}
The variance of the predictive distribution is an
\textit{instantaneous} parameter that varies over a database and
$J_{\mathrm{var}}(\{x\},W)$ must be an average over the predictive
distributions. The predictive distribution of each sample has a distinct
variance and the averaging method can be selected to have properties that are
advantageous for the specific application. As noted in section \ref{s:Problem}, the predictive distribution is a standard-form distribution $q(x|\alpha)$.

The predictive distribution $q(x|\alpha)$ is commonly a mixture distribution. Hence we must find an expression for  the variance of a mixture distribution. We first note that the mean of a mixture distribution is simply
  \begin{align}
  \mu =\E_q[X] = \sum_{k=1}^K \gamma_k \,\E_{q_k}[X] = \sum_{k=1}^K \gamma_k\, \mu_k,
  \end{align}
where $\E_q$ is expectation over $q$ and $q_k = \breve{q}(\cdot ; \mu_k , s_k)$, with $\breve{q}$ a mixture component. The variance of the mixture distribution is
  \begin{align}
    \E_q[(X-\E_q[X])^2] 
        &=   \sum_{k=1}^K \gamma_k  (\sigma_k^2 +\mu_k^2 - \mu^2) . \label{q:mixtvar}
  \end{align}
  
We now consider the specific case of a mixture of logistics in more
detail. The logistic distribution for component $k$ is:
\begin{align}
  \breve{q}(x; \mu_k, s_k) = \frac{\e^{-\frac{x-\mu_k}{s_k}}}{(1+\e^{-\frac{x-\mu_k}{s_k}})^2},
\end{align}
where $s_k$ is the scale and $\mu_k$ is an offset. It is easily seen that the logistic distribution is symmetric around $\mu $ and that, hence, $\mu$ is the distribution mean. The variance of the logistic distribution is
\begin{align}
  \E_{X\sim\breve{q}}[(X - \E_{X\sim\breve{q}}[X])^2]  = \frac{s^2 \pi^2}{3}. \label{q:logisvar}
\end{align}
We can now write down the variance of the mixture of logistics model by combining
\eqref{q:mixtvar} and \eqref{q:logisvar}: 
\begin{align}
\sigma^2_q  =  \sum_{k=1}^K \gamma_k  ( \frac{s_k^2 \pi^2}{3}  +\mu_k^2)   - (\sum_{k=1}^K \gamma_k \mu_k)^2 . \label{q:logismixtvar}
  \end{align}

\subsubsection{Regularization terms}
The most obvious approach for reducing the prediction variance is to use the
prediction variance \eqref{q:logismixtvar} directly as variance regularization in the
objective function \eqref{q:genObjFunc}: 
\begin{align}
  J_{\mathrm{var}}(\{x\}, W) = \E_{\mathrm{data}}[\sigma_q^2 ], \label{q:Jlinear}
\end{align}
where $\E_{\mathrm{data}}$ indicates averaging over the database. That is, we
encourage selection of weights $W$ of the network $\phi$ that minimize $\sigma_q^2$.

Straightforward optimization of \eqref{q:Jlinear} over a database may result in the prediction   variance being reduced mainly for signal regions where the conditional differential entropy
\eqref{q:condEnt} is large. The conditional differential entropy can be decomposed into the sum of a scale-independent term and a logarithmic scale (signal variance) dependency. For speech the scale-independent term is large for unvoiced segments while the scale-dependent term is large for voiced speech (as it is relatively loud).

For signals that have uniform overall signal variance, it may be desirable to encourage low predictive variance only for regions that have relatively low conditional differential entropy. (For speech that would correspond to encouraging low variance for voiced speech only.)
This can be accomplished by a monotonically increasing concave function of the predictive variance. The logarithm is particularly attractive for this purpose as it is invariant with scale: the
effect of a small variance getting smaller equals that of a large variance
getting smaller by the same proportion. We then have:
\begin{align}
  J_{\mathrm{var}}(\{x\}, W) = \E_{\mathrm{data}} [ \log  ( \sigma_q + a) ], \label{q:Jlog}
\end{align}
with $a$ providing a floor. 

\subsection{Baseline distribution approach}
\label{s:baseline}
For completeness we describe an alternative method for preventing the vanishing support problem of the Kullback-Leibler divergence by using a "baseline" distribution. To this  purpose consider a mixture distribution of the form
\begin{multline}
  q_{\mathrm{train}}(x_i;\phi) = \gamma_0 \,\breve{q}(x_i| \alpha_0) + \\\sum_{k=1}^K \gamma_k \breve{q}(x_i; \phi(x_{i-1},\cdots, x_{i-N}, \beta, W_k), \label{q:mixt+1}
\end{multline}
where the parameters \tcr{$\gamma_0$ and} $\alpha_0$ are set by the designer and where the first term is omitted during inference (the other terms must be renormalized by a \tcr{factor $\frac{1}{1-\gamma_0}$}).
By selecting $\alpha_0$ to provide an overly broad distribution, the distribution used for inference will be of low variance.
\section{System architecture}
\label{s:system}

In this section we describe the architecture of our coding scheme. The parameter settings of the scheme are provided in section \ref{s:Sysconfig}.

Let us consider an input signal with a sampling rate $S$ Hz. To avoid the need for modeling summed independent signals, the input is pre-processed with a real-time TasNet \cite{luo2019conv,sonning2020performance}. 

The encoder first converts the signal into a sequence of log mel spectra (e.g., \cite{o1987speech}). A set of subsequent log mel-spectra are stacked into a supervector that is subjected to a Karhunen-Lo\`eve transform (KLT) \tcr{that is optimized off-line}. The transformed stacked log mel spectra are encoded using split vector quantization with a small number of coefficients per split. No other information is encoded.

The decoder first decodes the bit stream into a sequence of quantized log mel spectra. These spectra form the input to the \textit{conditioning stack}, which consists of a set of 1D convolutional layers, all except the first with dilation. The output is a vector sequence \tcr{with a sampling rate equal to that of the mel spectra of the encoder and a dimensionality equal to the state of the GRU unit discussed below.}

The autoregressive network consists of a multi-band WaveGRU, which is based on gated recurring units (GRU) ) \cite{chung2014empirical}. For our $N$-band \textit{WaveGRU}, $N$ samples are generated simultaneously at an update rate of $S/N$ Hz, one sample for each frequency band. For each update, the state of the GRU network is projected onto an $N \times K\times 3$ dimensional space that defines $N$ parameter sets, each set corresponding to a mixture of logistics for a band. The value of a next signal sample for each band is then drawn by first selecting the mixture component (a logistics distribution) according to its probability and then drawing the sample from this logistic distribution by transforming a sample from a uniform distribution. 
For each set of $N$ samples a synthesis filter-bank produces $N$ subsequent time-domain samples, which results in an output with sampling rate $S$ Hz.

The input to the WaveGRU consists of the addition of an autoregressive and conditioning components. The autoregressive component is a projection of the last $N$ frequency-band samples projected onto a vector of the dimensionality of the WaveGRU state. The second component is the output of the conditioning stack (dimensionality of the WaveGRU state), repeated in time to obtain the correct sampling rate of $S/N$ Hz. 

The training of the GRU network and the conditioning stack is performed simultaneously using teacher forcing. That is, the past signal samples that are provided as input to the GRU are ground-truth signal samples. The objective function ) \label{q:genObjFunc}, combining log likelihood (cross entropy) and variance regularization, is used for each subsequent signal sample. For our implementation with variance regularization, we found the baseline distribution not to aid performance significantly, and it was omitted from the experiments.

\section{Experiments}
\label{s:experiments}
Our experiments had two goals. The first is to show the effect of predictive variance regularization and noise suppression. The second is to show that our contributions enable a practical system.

\subsection{System configuration}
\label{s:Sysconfig}
We tested eight systems, all variants based on a single baseline system \tcr{operating on} 16 kHz sampled signals. It is conditioned using a sequence of 160-dimensional log mel spectra computed from 80 ms windows at an update rate of 50 Hz. The system uses four frequency bands, each band sampled at 4 kHz. The conditioning stack consists of a single non-causal input layer (expanding from 160 channels to 512 channels), three dilated causal convolutional layers with kernel size two, and three upsampling transpose convolutional layers (kernel size two). \tcr{The overall algorithmic delay is 90 ms.} The conditioning outputs are tiled to match the GRU update frequency. The GRU state dimensionality is 1024, and eight mixture-of-logistics components are used for the predictive distribution per band.

The systems were trained from randomly initialized weights $W$ for 7.5 million steps, using a mini-batch size of 256.  The target signal was
from a combination of clean \cite{wsj0,librispeech} and noisy sources \cite{commonvoice}, including large proprietary TTS datasets. Additional noise was added from \cite{freesound}, with random SNR between $0$ and $40$~dB SNR. 

Table \ref{t:table} shows the combinations of coder attributes that were used. We briefly discuss each attribute. The variance regularization included refinements that further improved its performance: it was applied to the first two bands only and $\nu$ in \eqref{q:genObjFunc} was made proportional to a voicing score. The noise suppression system was a version of ConvTasNet \cite{sonning2020performance}. The weight pruning attribute was selected to enable implementation on consumer devices. For the three GRU matrices, we used block-diagonal matrices with 16 blocks, which uses 93\% fewer weights than a fully connected model. For other hidden layers, we applied iterative magnitude pruning to remove 92\% of the model weights \cite{zhu2017prune}. The pruning makes the codec \tcr{with TasNet} run reliably on a Pixel 2 phone in single-threaded mode.  The system was quantized with 120 bits per supervector, each supervector containing two log mel spectra, \tcr{for an overall rate of 3 kb/s}. The quantization was a two-dimensional vector-quantization of the KLT coefficients.

\begin{table}
\centering
\caption{Systems in test. \label{t:table}}
\begin{tabular}{|l|cccccccc|}
\hline
system label          & b & v & t & vt & q & qv & qt & qvt\\\hline
var. regularization    &   & \cm &   & \cm &   & \cm &   & \cm \\ 
TasNet noise supp. &   &   & \cm & \cm &   &   & \cm & \cm\\ 
3 kb/s quantization      &   &   &   &   & \cm & \cm & \cm & \cm \\  
93\% pruning           &   &   &   &   & \cm & \cm & \cm & \cm \\
\hline
\end{tabular}
\end{table}

\subsection{Testing procedure}
\label{s:TestingSetup}
To evaluate the absolute quality of the different systems on different SNRs a Mean Opinion Score (MOS) listening test was performed. \tcr{Except for data collection,} we followed the ITU-T P.800 \cite{P800} \tcr{(ACR)} recommendation. The data was collected using a crowd-sourcing platform with the requirements on listeners being native English speakers and using headphones. The evaluation dataset is composed of 30 samples from the Noisy VCTK dataset \cite{noisyvctk}: 15 clean and 15 augmented with additive noise at various SNRs (2.5, 7.5 and 12.5 dB). Each utterance for each system was rated about 200 times and the average and 95\% confidence interval were calculated per SNR.

\subsection{Results}
\label{s:results}

\begin{figure}
\centering
\includegraphics[width=0.36\textwidth]{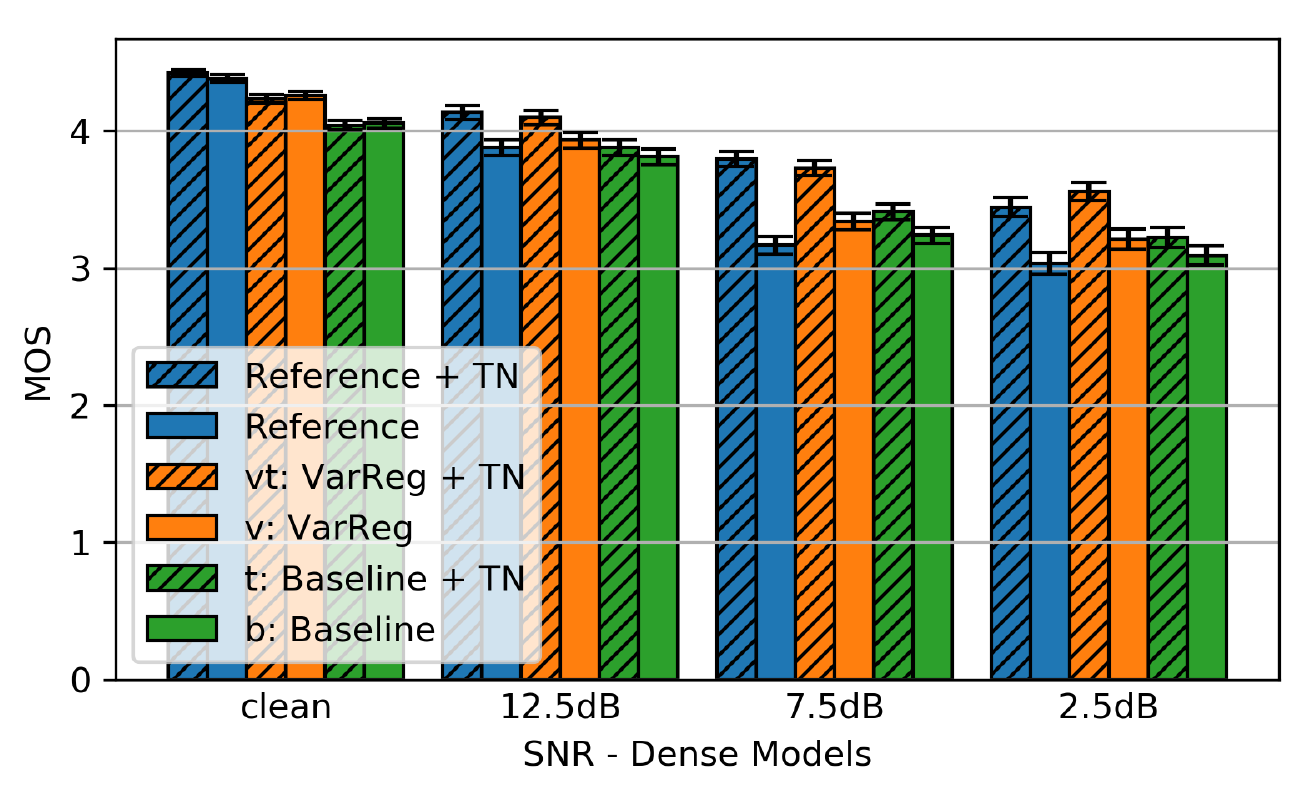}\vspace{-5mm}
\caption{Quality vs SNR for baseline, regularized, denoised, and regularized and denoised systems. Also shown are the quality of unprocessed and denoised only signals. \label{f:figFundamental}}
\end{figure}

\begin{figure}
\vspace{-2.5mm}
\centering
\includegraphics[width=0.367\textwidth]{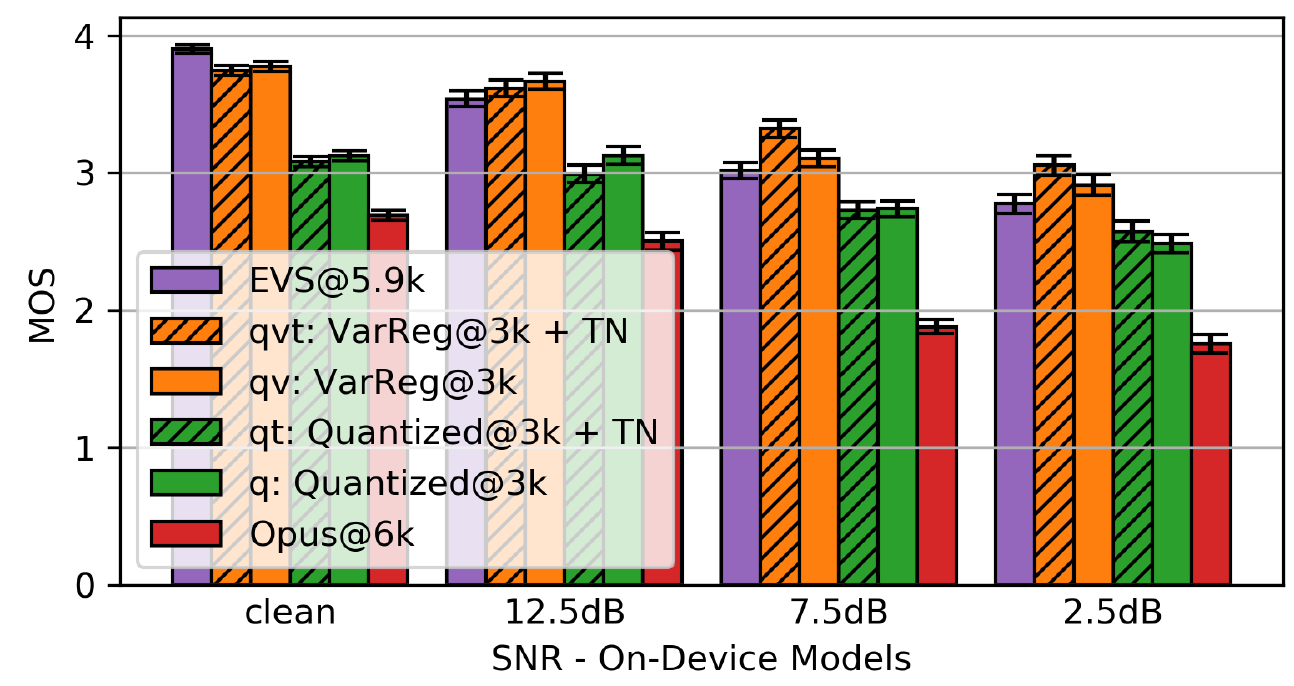}\vspace{-5mm}
\caption{Quality vs SNR for the pruned and quantized systems. \label{f:figQuant}} 
\vspace{-2mm}
\end{figure}

The quality for the systems of Table \ref{t:table} is shown in Figs. \ref{f:figFundamental} and \ref{f:figQuant}. The MOS with 95\% confidence intervals are given for four SNRs. 

Fig. \ref{f:figFundamental} displays the effect of predictive variance regularization and noise suppression \tcr{(TN)} without weight pruning and quantization. Predictive variance regularization results in a significant quality improvement and reduces the sensitivity to noise in the input signal. Noise suppression aids performance when noise is present.

Fig. \ref{f:figQuant} shows the quality for pruned and quantized systems. For this case, the improvement due to variance regularization is particularly large for clean signals. The effect of noise suppression \tcr{(TN)} varies in an unexpected manner with SNR. This likely results from an interaction between noise suppression and quantization. It may be related to noise suppression reducing signal variability and quantization reducing noise on its own.

As a reference, Fig. \ref{f:figQuant} provides the performance of the Opus codec \cite{Opus} operating at 6~kb/s and the EVS codec \cite{dietz2015evs} operating at 5.9~kb/s \tcr{(for fairness with disabled DTX)}. It is seen that the proposed fully practical 3 kb/s WaveGRU coder performs significantly better than Opus at 6~kb/s and similarly to EVS operating at 5.9~kb/s.

\section{Conclusion}
\label{s:conclusion}
We have developed a robust speech codec using neural-network based signal synthesis that encodes speech at 3 kb/s. Our system is \tcr{suitable for, for example, low-rate video calls, and fits in consumer devices as evidenced by our implementation} running on a wide range of mobile phones including the Pixel 2. Our experiments show that its quality is similar or better than state-of-the-art conventional codecs operating at double the rate. Our main contribution is that we addressed the impact of variability and distortion inherent in real-world input to practical speech codecs. We identified as causes for poor performance \textit{i}) the inherent emphasis of outliers by the maximum likelihood criterion and \textit{ii}) the difficulty of modeling a sum of multiple independent sources. We resolved these problems \tcr{with predictive variance regularization and noise suppression}.

\bibliographystyle{IEEEtran}
\bibliography{ref_learning,ref_coding,ref_learning02,ref_misc}

\begin{thebibliography}{10}
\providecommand{\url}[1]{#1}
\csname url@samestyle\endcsname
\providecommand{\newblock}{\relax}
\providecommand{\bibinfo}[2]{#2}
\providecommand{\BIBentrySTDinterwordspacing}{\spaceskip=0pt\relax}
\providecommand{\BIBentryALTinterwordstretchfactor}{4}
\providecommand{\BIBentryALTinterwordspacing}{\spaceskip=\fontdimen2\font plus
\BIBentryALTinterwordstretchfactor\fontdimen3\font minus
  \fontdimen4\font\relax}
\providecommand{\BIBforeignlanguage}[2]{{%
\expandafter\ifx\csname l@#1\endcsname\relax
\typeout{** WARNING: IEEEtran.bst: No hyphenation pattern has been}%
\typeout{** loaded for the language `#1'. Using the pattern for}%
\typeout{** the default language instead.}%
\else
\language=\csname l@#1\endcsname
\fi
#2}}
\providecommand{\BIBdecl}{\relax}
\BIBdecl

\bibitem{oord2016wavenet}
A.~v.~d. Oord, S.~Dieleman, H.~Zen, K.~Simonyan, O.~Vinyals, A.~Graves,
  N.~Kalchbrenner, A.~Senior, and K.~Kavukcuoglu, ``Wavenet: A generative model
  for raw audio,'' \emph{arXiv preprint arXiv:1609.03499}, 2016.

\bibitem{kalchbrenner2018efficient}
N.~Kalchbrenner, E.~Elsen, K.~Simonyan, S.~Noury, N.~Casagrande, E.~Lockhart,
  F.~Stimberg, A.~v.~d. Oord, S.~Dieleman, and K.~Kavukcuoglu, ``Efficient
  neural audio synthesis,'' \emph{arXiv preprint arXiv:1802.08435}, 2018.

\bibitem{oord2018parallel}
A.~v.~d. Oord, Y.~Li, I.~Babuschkin, K.~Simonyan, O.~Vinyals, K.~Kavukcuoglu,
  G.~Driessche, E.~Lockhart, L.~Cobo, F.~Stimberg \emph{et~al.}, ``Parallel
  {WaveNet}: Fast high-fidelity speech synthesis,'' in \emph{International
  conference on machine learning}.\hskip 1em plus 0.5em minus 0.4em\relax PMLR,
  2018, pp. 3918--3926.

\bibitem{prenger2019waveglow}
R.~Prenger, R.~Valle, and B.~Catanzaro, ``Waveglow: A flow-based generative
  network for speech synthesis,'' in \emph{2019 IEEE Int. Conf. Acoust Speech
  Signal Processing (ICASSP)}.\hskip 1em plus 0.5em minus 0.4em\relax IEEE,
  2019, pp. 3617--3621.

\bibitem{donahue2018adversarial}
C.~Donahue, J.~McAuley, and M.~Puckette, ``Adversarial audio synthesis,''
  \emph{arXiv preprint arXiv:1802.04208}, 2018.

\bibitem{engel2019gansynth}
J.~Engel, K.~K. Agrawal, S.~Chen, I.~Gulrajani, C.~Donahue, and A.~Roberts,
  ``{GANSynth}: Adversarial neural audio synthesis,'' \emph{arXiv preprint
  arXiv:1902.08710}, 2019.

\bibitem{kleijn2018wavenet}
W.~B. Kleijn, F.~S. Lim, A.~Luebs, J.~Skoglund, F.~Stimberg, Q.~Wang, and T.~C.
  Walters, ``{WaveNet} based low rate speech coding,'' in \emph{2018 IEEE Int.
  Conf. Acoust Speech Signal Processing (ICASSP)}.\hskip 1em plus 0.5em minus
  0.4em\relax IEEE, 2018, pp. 676--680.

\bibitem{klejsa2019high}
J.~Klejsa, P.~Hedelin, C.~Zhou, R.~Fejgin, and L.~Villemoes, ``High-quality
  speech coding with sample {RNN},'' in \emph{2019 IEEE Int. Conf. Acoust
  Speech Signal Processing (ICASSP)}.\hskip 1em plus 0.5em minus 0.4em\relax
  IEEE, 2019, pp. 7155--7159.

\bibitem{garbacea2019low}
C.~G{\^a}rbacea, A.~van~den Oord, Y.~Li, F.~S. Lim, A.~Luebs, O.~Vinyals, and
  T.~C. Walters, ``Low bit-rate speech coding with {VQ-VAE} and a {WaveNet}
  decoder,'' in \emph{2019 IEEE Int. Conf. Acoust Speech Signal Processing
  (ICASSP)}.\hskip 1em plus 0.5em minus 0.4em\relax IEEE, 2019, pp. 735--739.

\bibitem{valin2019real}
J.-M. Valin and J.~Skoglund, ``{A Real-Time Wideband Neural Vocoder at 1.6kb/s
  Using LPCNet},'' in \emph{Proc. Interspeech 2019}, 2019, pp. 3406--3410.

\bibitem{lim2020robust}
F.~S. Lim, W.~B. Kleijn, M.~Chinen, and J.~Skoglund, ``Robust low rate speech
  coding based on cloned networks and wavenet,'' in \emph{2020 IEEE Int. Conf.
  Acoust Speech Signal Processing (ICASSP)}.\hskip 1em plus 0.5em minus
  0.4em\relax IEEE, 2020, pp. 6769--6773.

\bibitem{fejgin2020source}
R.~Fejgin, J.~Klejsa, L.~Villemoes, and C.~Zhou, ``Source coding of audio
  signals with a generative model,'' in \emph{2020 IEEE Int. Conf. Acoust
  Speech Signal Processing (ICASSP)}.\hskip 1em plus 0.5em minus 0.4em\relax
  IEEE, 2020, pp. 341--345.

\bibitem{granger1976time}
C.~W.~J. Granger and M.~J. Morris, ``Time series modelling and
  interpretation,'' \emph{Journal of the Royal Statistical Society: Series A
  (General)}, vol. 139, no.~2, pp. 246--257, 1976.

\bibitem{goodfellow2014generative}
I.~Goodfellow, J.~Pouget-Abadie, M.~Mirza, B.~Xu, D.~Warde-Farley, S.~Ozair,
  A.~Courville, and Y.~Bengio, ``Generative adversarial nets,'' in
  \emph{Advances in neural information processing systems}, 2014, pp.
  2672--2680.

\bibitem{nowozin2016f}
S.~Nowozin, B.~Cseke, and R.~Tomioka, ``{f-GAN}: Training generative neural
  samplers using variational divergence minimization,'' in \emph{Advances in
  neural information processing systems}, 2016, pp. 271--279.

\bibitem{arjovsky2017wasserstein}
M.~Arjovsky, S.~Chintala, and L.~Bottou, ``Wasserstein {GAN},'' \emph{arXiv
  preprint arXiv:1701.07875}, 2017.

\bibitem{li2015generative}
Y.~Li, K.~Swersky, and R.~Zemel, ``Generative moment matching networks,'' in
  \emph{International Conference on Machine Learning}, 2015, pp. 1718--1727.

\bibitem{li2017mmd}
C.-L. Li, W.-C. Chang, Y.~Cheng, Y.~Yang, and B.~P{\'o}czos, ``{MMD} {GAN}:
  Towards deeper understanding of moment matching network,'' in \emph{Advances
  in Neural Information Processing Systems}, 2017, pp. 2203--2213.

\bibitem{kim2018flowavenet}
S.~Kim, S.-g. Lee, J.~Song, J.~Kim, and S.~Yoon, ``{FloWaveNet}: A generative
  flow for raw audio,'' \emph{arXiv preprint arXiv:1811.02155}, 2018.

\bibitem{luo2019conv}
Y.~Luo and N.~Mesgarani, ``{Conv-TasNet}: Surpassing ideal time--frequency
  magnitude masking for speech separation,'' \emph{IEEE/ACM Transactions on
  Audio, Speech, and Language Processing}, vol.~27, no.~8, pp. 1256--1266,
  2019.

\bibitem{sonning2020performance}
S.~Sonning, C.~Sch{\"u}ldt, H.~Erdogan, and S.~Wisdom, ``Performance study of a
  convolutional time-domain audio separation network for real-time speech
  denoising,'' in \emph{2020 IEEE Int. Conf. Acoust Speech Signal Processing
  (ICASSP)}.\hskip 1em plus 0.5em minus 0.4em\relax IEEE, 2020, pp. 831--835.

\bibitem{o1987speech}
D.~{O'Shaughnessy}, \emph{Speech Communications: Human And Machine
  (IEEE)}.\hskip 1em plus 0.5em minus 0.4em\relax Universities press, 1987.

\bibitem{chung2014empirical}
J.~Chung, C.~Gulcehre, K.~Cho, and Y.~Bengio, ``Empirical evaluation of gated
  recurrent neural networks on sequence modeling,'' \emph{arXiv preprint
  arXiv:1412.3555}, 2014.

\bibitem{wsj0}
\BIBentryALTinterwordspacing
J.~S. Garofolo, D.~Graff, D.~Paul, and D.~Pallett, ``{CSR-I (WSJ0) Other},''
  Harvard Dataverse, Tech. Rep., 2016. [Online]. Available:
  \url{https://doi.org/10.7910/DVN/ZVU9HF}
\BIBentrySTDinterwordspacing

\bibitem{librispeech}
V.~Panayotov, G.~Chen, D.~Povey, and S.~Khudanpur, ``Librispeech: an {ASR}
  corpus based on public domain audio books,'' in \emph{2015 IEEE International
  Conference on Acoustics, Speech and Signal Processing (ICASSP)}.\hskip 1em
  plus 0.5em minus 0.4em\relax IEEE, 2015, pp. 5206--5210.

\bibitem{commonvoice}
R.~{Ardila et al.}, ``Common voice: A massively-multilingual speech corpus,''
  \emph{arXiv preprint arXiv:1912.06670}, 2019.

\bibitem{freesound}
E.~Fonseca, J.~Pons, X.~Favory, F.~Font, D.~Bogdanov, A.~Ferraro, S.~Oramas,
  A.~Porter, and X.~Serra, ``Freesound datasets: a platform for the creation of
  open audio datasets,'' in \emph{Proc. 18th Int. Society Music Information
  Retrieval Conference (ISMIR 2017)}, Suzhou, China, 2017, pp. 486--493.

\bibitem{zhu2017prune}
M.~Zhu and S.~Gupta, ``To prune, or not to prune: exploring the efficacy of
  pruning for model compression,'' \emph{arXiv preprint arXiv:1710.01878},
  2017.

\bibitem{P800}
\emph{Recommendation ITU-T P.800 Methods for subjective determination of
  transmission quality}, ITU-T Std., Aug 1996.

\bibitem{noisyvctk}
C.~Valentini-Botinhao, ``Noisy speech database for training speech enhancement
  algorithms and tts models,'' University of Edinburgh. School of Informatics.
  Centre for Speech Technology Research (CSTR), Tech. Rep., 2016.

\bibitem{Opus}
J.~M. Valin, K.~Vos, and T.~Terriberry, \emph{Definition of the Opus Audio
  Codec}, IETF Std., Sept 2012, {RfC}: 6717.

\bibitem{dietz2015evs}
M.~{Dietz $et\,\,al$}, ``Overview of the {EVS} codec architecture,'' in
  \emph{2015 IEEE Int. Conf. Acoust Speech Signal Processing (ICASSP)}.\hskip
  1em plus 0.5em minus 0.4em\relax IEEE, 2015, pp. 5698--5702.

\end{thebibliography}
\end{document}